\documentstyle[aps,prl,twocolumn,graphics]{revtex}
\newcommand{\ra}{\rangle}
\newcommand{\la}{\langle}

\begin{document}
\draft
\wideabs{
\title{\bf Super-poissonian photon statistics and correlations between pump 
and probe fields in Electromagnetically Induced Transparency}
\author{C. L. Garrido Alzar, L. S. Cruz, J. G. Aguirre G\'omez, M. 
Fran\c{c}a Santos,\cite{byline} and P. Nussenzveig}
\address{Instituto de F\'\i sica, Universidade de S\~ao Paulo, 
Caixa Postal 66318, CEP 05315-970, S\~ao Paulo, SP, Brazil.}
\date{\today}
\maketitle

\begin{abstract} 

We have measured the photon statistics of pump and probe beams after 
interaction with Rb atoms in a situation of Electromagnetically Induced 
Transparency. Both fields present super-poissonian statistics and 
their intensities become correlated, in good qualitative agreement with 
theoretical predictions in which both fields are treated quantum-mechanically. 
The intensity correlations measured are a first step towards the observation 
of entanglement between the fields. 

\end{abstract}

\pacs{PACS numbers: 42.50.-p, 42.50.Ar, 42.50.Ct, 42.50.Gy}
}

Electromagnetically Induced Transparency (EIT)\cite{harris} is an interference 
effect 
that can be observed when three-level atoms interact simultaneously with 
two lasers. The recent interest in these systems has been driven by 
observations of very slow light pulse propagation in EIT media~\cite{hau,kash,budker} and 
even of light storage~\cite{hau2,lukin}. Even though it is normally viewed as a quantum 
interference effect, EIT has a very simple classical counterpart~\cite{ajp,hemmprent88}. 
The question 
of whether there exist intrinsic quantum effects (with no classical analog) 
in EIT naturally arises in this context. It is even more important in connection 
with suggested applications in the field of quantum 
information~\cite{lukimam,lukyelfleisch}. Quantum field properties in EIT 
have been theoretically investigated. Phase-noise squeezing was predicted by 
Fleischhauer and co-workers~\cite{fleischscully}. Agarwal~\cite{agarwal} found 
matched photon-statistics for two classical and 
two quantum fields interacting with three-level atoms, in a situation of 
Coherent Population Trapping. Jain~\cite{jain} extended this work, and 
predicted excess-noise correlations in EIT. In all the 
above~\cite{lukimam,lukyelfleisch,fleischscully,agarwal,jain}, at least 
one of the fields was treated classically. As we will see below, it is 
our belief that such an approximation is not valid in the EIT situation. 

In this Letter we present the first experimental investigation (to our knowledge) 
of field fluctuations and correlations in EIT. We have performed photon statistics 
measurements of both pump and probe fields, as a function 
of the probe field detuning. As predicted by our theoretical treatment, 
both fields, treated quantum-mechanically, have super-poissonian statistics. 
Furthermore, they are coupled by their interaction with the atoms and thus their 
intensities become correlated. 

We begin by describing our theoretical model for three-level atoms in the 
$\Lambda$-configuration (ground states $|1\rangle$ and $|2\rangle$, 
and excited state $|0\rangle$) interacting with two quantum modes of the 
electromagnetic field. In order to simulate the interaction 
of the atoms with two propagating fields, we use the input-output 
formalism~\cite{hilico,courty}, and consider the interaction between the 
two fields and the atoms in a ring cavity~\cite{figprocessus}, with external 
input fields. For the intracavity field operators we used plane 
wave and quasi-monochromatic approximations. The interaction hamiltonian is 
obtained with the usual dipole and rotating-wave approximations

\begin{equation}
\hat{H}_{\mbox{\scriptsize int}} = \hbar g_1 \hat{S}^+_1(t) \hat{A}_1(t) 
+ \hbar g_2 \hat{S}^+_2(t) \hat{A}_2(t) + \mbox{h.c.}\;,
\label{hamilt}
\end{equation}

\noindent 
where $\hat{A}_1$ ($\hat{A}_2$) is the annihilation operator for 
intracavity field 1 - pump (intracavity field 2 - probe), $g_1$ ($g_2$) 
is the atom -- field 1 (field 2) coupling strength, and $\hat{S}^+_1$ 
($\hat{S}^+_2$) the atomic polarization on the transition 
$|1\rangle \leftrightarrow |0\rangle$ ($|2\rangle \leftrightarrow 
|0\rangle$).

From the Heisenberg equations of motion, we derive quantum Langevin 
equations~\cite{Processus} for the system operators. In a matrix form, we 
find~\cite{tobe} 

\begin{equation}
 \frac{d\widehat{\bar{X}}(t)}{dt}=-\hat{\mbox{\bf A}}(t)~\widehat{\bar{X}}(t) +
 {\mbox{\bf D}}\widehat{\bar{\cal F}}(t) \;.
\label{eq1}
\end{equation}
The vector operator $\widehat{\bar{X}}(t)$ has as its components the operators 
giving the atomic inversions and polarizations (and their hermitian conjugates) 
corresponding to the two transitions, the coherences between the two 
ground states and the annihilation and creation operators of both fields. 
The vector of the Langevin forces acting on the 
system is given by $\widehat{\bar{\cal F}}(t)$, and {\bf D} is the 
diffusion matrix. The matrix $\hat{\mbox{\bf A}}(t)$, in the steady state, 
will represent the drift matrix. Eq.~(\ref{eq1}) is a compact form of writing 
twelve coupled differential equations for the operators. 

We are interested in studying the fluctuations of the field operators. This 
is done by linearizing the operators in Eq.~(\ref{eq1}) around their 
stationary values. A new set of twelve coupled differential equations is 
obtained. It is interesting to examine the equations for the fluctuations of 
one field and for the corresponding atomic polarization: 

\begin{equation} 
 \frac{d\delta \hat{A}_1}{dt}=-\left(\frac{\gamma}{2} + i 
\Delta_{c1}\right)\delta \hat{A}_1 - i \frac{g_1}{\tau} \delta \hat{S}^-_1 
+ \sqrt{\frac{\gamma}{\tau}} \delta\hat{A}_{1in} \;,
\label{eq:fluctA1} 
\end{equation} 

\begin{eqnarray} 
 \frac{d\delta \hat{S}^-_1}{dt}& = & -\left(\frac{\Gamma_1}{4} - i 
\delta_{L1}\right)\delta \hat{S}^-_1 + i g_1 w_1 \delta\hat{A}_1 + i g_1 \alpha_1 
\delta\hat{W}_1 \nonumber \\
 & & - i g_2 s^*_{12} \delta\hat{A}_2 
- i g_2 \alpha_2 \delta\hat{S}^+_{12} + \hat{F}_{S1} \;.
\label{eq:fluctS1} 
\end{eqnarray} 

\noindent
Here we define $\gamma$ as the cavity 
linewidth, $\Delta_{c1}$ 
cavity detuning for field 1, $\tau$ cavity length divided by the speed 
of light, $\hat{A}_{1in}$ the annihilation operator for the input field 1, $\Gamma_1$ 
spontaneous emission rate from $|0\rangle \rightarrow |1\rangle$, $\delta_{L1}$ 
detuning between field 1 and the corresponding atomic transition, $w_1$ steady-state 
inversion between states $|0\rangle$ and $|1\rangle$, $\alpha_1$ ($\alpha_2$) 
steady-state amplitude of field 1 (field 2), $\hat{W}_1$ inversion (operator) 
between states $|0\rangle$ and $|1\rangle$, $s^*_{12}$ steady-state coherence 
between ground states $|1\rangle$ and $|2\rangle$, $\hat{S}^+_{12}$ coherence 
operator, $\hat{F}_{S1}$ Langevin fluctuation force. The notation 
$\delta \hat{S}^-_1$ means fluctuations of the corresponding operator. 

It is clear from Eqs.~(\ref{eq:fluctA1}) and (\ref{eq:fluctS1}) that 
the fluctuations in field 1 are not only determined by input field fluctuations 
but also by atomic fluctuations and by fluctuations in field 2. Moreover, in Eq.~(\ref{eq:fluctS1}) we notice that noise correlations between the
fields will arise, conditioned to the existence of a coherence between the 
ground states, $s^*_{12}$, only important on EIT resonance. 

The terms $i g_1 w_1 \delta\hat{A}_1$, $i g_1 \alpha_1 
\delta\hat{W}_1$, $i g_2 s^*_{12} \delta\hat{A}_2$, 
$i g_2 \alpha_2 \delta\hat{S}^+_{12}$ in Eq.~(\ref{eq:fluctS1}) are all 
of the same order. This means that we can not neglect the fluctuations 
of either field with respect to its average (steady-state value). It is 
not valid to treat one field as classical and the other as quantized. 
This conclusion also applies to the work by Agarwal~\cite{agarwal} 
and Jain~\cite{jain} who consider four independent fields (two classical 
and two quantized) interacting with coherent population trapped (CPT) 
atoms. The fields used to produce CPT also have inherent quantum 
fluctuations, which lead to atomic fluctuations and, therefore, should 
not be neglected. 

From the twelve coupled differential equations for the fluctuations, we 
calculate the spectral matrix {\bf S}$(\Omega)$~\cite{walls,collet} 
for the noise, which yields the fluctuations of the output cavity fields, 
after Fourier transform of the input-output relations. 

The use of a cavity to simulate an EIT experiment with cw propagating 
fields (no cavity) is adequate as long as we restrict ourselves to 
fluctuations around the steady state and do not investigate transient 
effects. We must also consider the cavity large enough so that the atoms 
experience neither significant changes in their spontaneous emission rates 
nor collective effects.

\begin{figure} 
\centering \resizebox{7.5cm}{!}{\includegraphics*{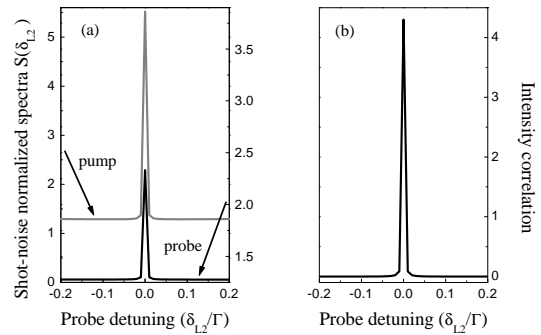}}
\caption{Theoretical predictions. (a) Super-poissonian photon 
statistics for both fields, as a function of probe detuning. The 
left vertical axis is for the pump (gray) and right axis for the probe (black). 
(b) Intensity correlations as a function of probe detuning (in units of 
$\Gamma$).} 
\label{fig:teoria} 
\end{figure}

The noise spectra calculated for probe and pump amplitude quadratures and their 
correlations are plotted in Fig.~\ref{fig:teoria}, as a function of the probe frequency. 
These spectra were calculated for an analysis frequency $\Omega = \Gamma/6\pi$ 
(consistent with the experiment), where $\Gamma$ is the total spontaneous emission 
rate from the excited state. We took equal coupling constants for both transitions, 
pump strictly on resonance, and a ratio of intensities [$|\alpha_1/\alpha_2|^2 = 9$ 
in Fig.~\ref{fig:teoria}~(a)] such that there is a very narrow and deep EIT resonance. 
For the correlations in Fig.~\ref{fig:teoria}~(b), both fields have equal intensities. 
The number of atoms was of the order of $10^4$, which is sufficiently large to justify 
the linearization method used. 

We find super-poissonian statistics for both fields, and intensity correlations 
peaked at the EIT resonance. Physically we understand this behavior as follows. 
The coupling of the fields with the atoms introduces a spreading in the photon 
distributions associated to each field. The atoms act as ``beamsplitters'', 
redistributing photons between both modes, in such a way that the mean 
numbers do not vary appreciably but the variances are modified. Taking 
absorption into account, we expect this effect to be reduced, since absorption 
introduces randomness in the photon redistribution process. We therefore 
expect to observe a maximum effect for minimum absorption, which corresponds to 
the EIT resonance. The predicted correlations also depend on the presence 
of absorption. In the EIT situation, the transmission of one field depends on 
the (fluctuating) intensity of the other field (cross Kerr effect), leading to 
intensity correlations. The experimental results presented below are in very good 
qualitative agreement with these predictions. 

\begin{figure}
\centering \resizebox{7.5cm}{!}{\includegraphics*{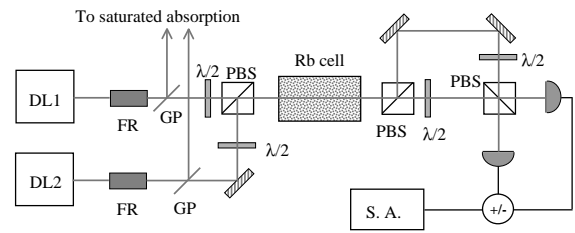}}
\caption{Sketch of the experimental setup. DL1 and DL2: extended-cavity 
diode lasers; FR: Faraday rotator; GP: glass plate; PBS: polarizing 
beamsplitter cube; S.A.: spectrum analyzer.} 
\label{fig:expset}
\end{figure}

Our experimental setup is sketched in Fig.~\ref{fig:expset}. Measurements 
were performed in a 5 cm long room-temperature Rb vapor cell. 
Two independent extended-cavity diode lasers (ECDLs) provide the pump and 
probe fields. A small portion of each beam 
is extracted and sent to auxiliary saturated absorption cells, 
used as frequency references. The pump beam is tuned to the 
$^{85}$Rb $|5S_{1/2}, F = 3\ra \rightarrow |5P_{3/2}, F' = 3\ra$ 
transition, while the probe beam is scanned across the 
Doppler-broadened $|5S_{1/2}, F = 2\ra \rightarrow |5P_{3/2}, 
F' = 1, 2, {\mbox{or }} 3\ra$ line. The two beams have orthogonal 
polarizations and are combined by means of a polarizing 
beamsplitter cube (PBS). At the output of the cell we can 
separate the beams again with another PBS. We then have the 
option of detecting either beam with a balanced detection 
setup~\cite{homodyne}. Half-wave plates also enable us to 
recombine the two fields and measure the photon statistics 
of the sum and difference intensities of the two beams. From 
these we can extract the intensity correlations. 

\begin{figure}
\centering \resizebox{8cm}{!}{\includegraphics*{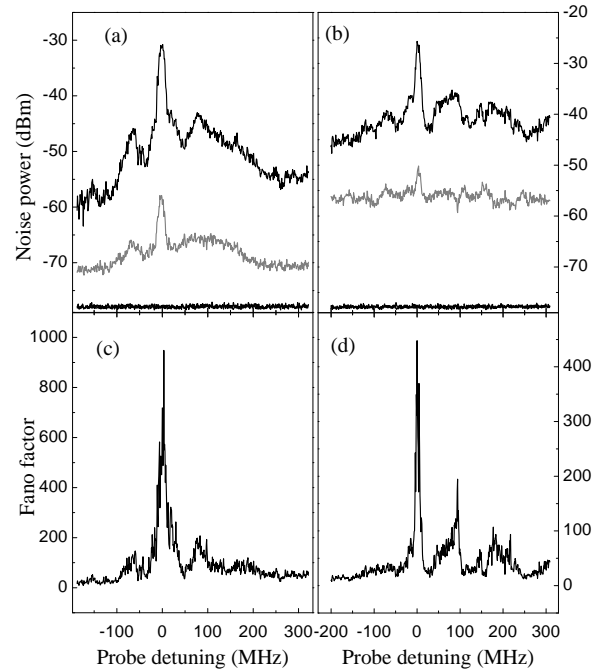}}
\caption{(a) and (b) Noise spectra of the probe and pump beams, respectively, 
as a function of the probe frequency and corresponding shot (in gray) and 
electronic noise (lower traces). (c) and (d) Fano factors deduced from (a) 
and (b).} 
\label{fig:probstat1}
\end{figure}

The usual EIT signals are observed by sending only the probe 
beam into the detection region and measuring the DC (average) 
intensity. The intensity transmitted through the 
cell has a narrow peak as a function of the probe frequency. The 
width is much narrower than the Doppler width, which is a signature 
of the interference effect. Next, we measure the photon statistics (intensity 
fluctuations) for the probe beam, yielding the signal presented in 
Fig.~\ref{fig:probstat1} (a). 
For the range of frequencies spanned, the photon statistics is always 
super-poissonian, but it presents a sharp peak (almost 20 dB) corresponding 
to the EIT resonance. In this measurement, the initial photon 
statistics of the probe beam is poissonian (for the frequency of 
analysis chosen, the pump initially has super-poissonian statistics, but 
the theory predicts this effect even when both fields have 
poissonian statistics). The pump photon statistics presents similar 
behavior, as seen in Fig.~\ref{fig:probstat1} (b). 
In Fig.~\ref{fig:probstat1} (c) and (d) we present the corresponding 
Fano factors for both probe and pump fields, respectively. The Fano 
factor is given by the ratio of the intensity fluctuations to the average 
intensity (shot noise level), and we plot it on a linear scale. In these 
measurements, the probe and pump intensities were 14.5~mW/cm$^2$ and 
63.8~mW/cm$^2$, respectively. The observed behavior agrees very well with 
the theoretical predictions of Fig.~\ref{fig:teoria}. 

The noise spectra were recorded using a spectrum analyzer (HP 8560 E) in 
the zero span mode ($RBW=300$~kHz, $VBW=3$~kHz), with a center (analysis) 
frequency of 2~MHz. This frequency is in a window such that the probe 
laser has poissonian photon statistics and the electronic noise is safely 
below shot noise (more than 5~dB). The pump laser's noise is $\sim 8$~dB 
above shot noise (and shows little variation with the analysis frequency). 
The analyzer 
is triggered by the same signal used to scan the probe frequency. The 
photodetectors are EG\&G FND-100 with a nominal efficiency of $\sim 70\%$. 
We checked that none of the measurements is influenced by saturation in the 
detection. This is done by introducing neutral density filters before 
detection and observing that the photon statistics becomes poissonian, with 
a linear dependence on absorption. 

For typical EIT signals, in the presence of a strong pump field, 
the shot noise level also peaks, as expected, 
following the increasing DC intensity on EIT resonance. However, if we 
lower the pump power, this effect tends to disappear, since 
incoherent effects then dominate over the coherent pumping on 
the transition $|5S_{1/2}, F = 3\ra \rightarrow |5P_{3/2}, F' = 3\ra$. 
The probe DC signal, and consequently the shot noise level, no longer 
show any evidence of a coherent effect in the atomic medium for 
pump intensities lower than 0.45~mW/cm$^2$, for a probe intensity of 
13.3~mW/cm$^2$ (it is the ratio of pump to probe intensities that 
is relevant). On the other hand, by looking at the 
photon statistics, the coherent effect is still clearly identifiable. 

\begin{figure}
\centering \resizebox{7cm}{!}{\includegraphics*{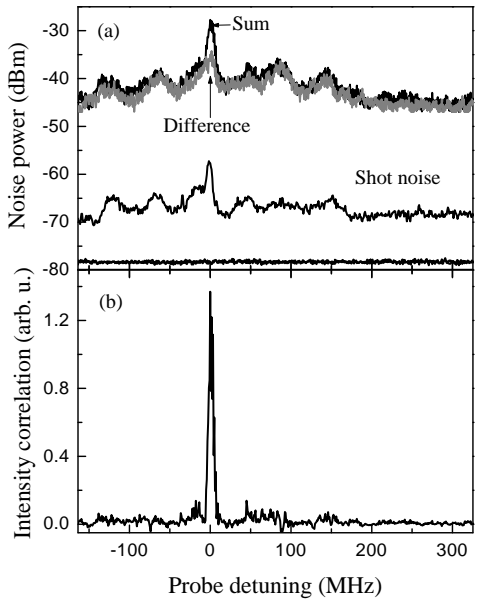}}
\caption{(a) Measurements of the noise in the sum (black) and difference (gray) 
intensities 
of pump and probe beams. The difference observed on EIT resonance 
is a measurement of the intensity correlation (b), on a linear scale, 
created between both 
fields. Pump and probe intensities: 7.8~mW/cm$^2$ and 7.5~mW/cm$^2$, 
respectively.} 
\label{fig:correl}
\end{figure}

We have also measured the noise properties of the sum and difference 
beam intensities. This is done by sending each beam to 
one of the detectors. By 
subtracting the intensity-difference noise from the intensity-sum, we 
observe a clear correlation on the EIT resonance. The corresponding 
shot-noise level is measured by mixing the two beams so that each 
detector receives half of each beam. The results are presented in 
Fig.~\ref{fig:correl}. We observe $\sim 7$~dB splitting between the sum and 
difference fluctuations. For photon statistics, this corresponds to 
a correlation term $2C=4\{\la n_1n_2 \ra - \la n_1 \ra \la n_2 \ra \}$. 

In summary, we have measured the photon statistics of probe and 
pump beams after interaction with an atomic medium, in a situation 
of Electromagnetically Induced Transparency. We observe super-poissonian 
statistics, peaked on EIT resonance. We have also observed intensity 
correlations between the two initially independent fields, as a result 
of their interaction with the atoms. The experimental 
result features agree very well with our theoretical 
predictions. The balanced detection scheme leads to enhanced sensitivity 
for detecting coherent effects in the atomic medium. The importance of 
the correlations predicted and observed is two-fold. On the one hand, this 
is a very strong indication that, when investigating quantum properties 
of one of the fields in a coherent situation (such as EIT), one can not 
treat either field as classical, as done by several authors. The coherent 
exchange of photons between the fields, which is at the very origin of the 
EIT effect, introduces correlations between their fluctuations. Another 
important issue is the nature of the correlations between both fields. 
Quantum correlations (entanglement) between intense fields can be used 
in the context of quantum information. This would be the dual of the 
experiment by Julsgaard and co-workers~\cite{polzik}, in which correlations 
between atoms in two different vapor cells were created using light. 
In the present experiment, we have made the first step in this direction 
by demonstrating the existence of correlations. However, in order to 
distinguish between classical and quantum correlations, we will need an 
improved experimental setup. A new calculation shows that entanglement 
between pump and probe fields {\it can be} produced in the EIT situation, 
and its observation will require measurements of quadrature fluctuations 
of both fields~\cite{entangtobe}.

We thank Profs. L. M. Narducci and H. M. Nussenzveig for critically reading 
the manuscript, and S. Simionatto for technical assistance. One of us (M.F.S.) 
would like to thank Prof. S. Salinas for his hospitality in S\~ao Paulo. This 
research was funded by FAPESP. Additional support was 
provided by the Brazilian agencies CAPES and CNPq.


\begin{thebibliography}{99}
\bibitem{figprocessus} This approximation is quite common. See, for example, 
ref.~\cite{Processus}, pg. 410, Fig. 2.
\end{thebibliography}
\end{document}